\documentclass[]{aa}
\usepackage{graphicx,color}
\usepackage{amsfonts,amssymb,latexsym}
\usepackage{epstopdf}
\usepackage{xfrac}
\usepackage{natbib}

\def\ga{\,\,\raise0.14em\hbox{$>$}\kern-0.76em\lower0.28em\hbox{$\sim$}\,\,}
\def\la{\,\,\raise0.14em\hbox{$<$}\kern-0.76em\lower0.28em\hbox{$\sim$}\,\,}
\def\myr{$M_{\odot}$\,yr$^{-1}$}
\def\Msun{$\,M_{\odot}$}
\def\msun{\,M_{\odot}}
\def\Rsun{$R_{\odot}$}
\def\chem#1{$^{#1}$}
\def\reac#1#2#3#4#5#6{$\mathrm{\, ^{#2}\kern-0.8pt{#1}\, ({#3}\, ,{#4})\, {}^{#6}\kern-0.8pt{#5}\, }$}
\def\Mec{$M_\mathrm{EC}$}

\def\gcm{g\,cm$^{-3}$}

\def\be{\begin{equation}}
\def\ee{\end{equation}}
\def\LS#1#2{{\bf \color{black} #1}{}}

\begin{document}

\title{Case A and B evolution towards electron capture supernova}

\author{L. Siess \inst{1,2} \and U. Lebreuilly\inst{1} }

\offprints{L. Siess}

\institute{
Institut d'Astronomie et d'Astrophysique, Universit\'e Libre de Bruxelles
(ULB),  CP 226, B-1050 Brussels, Belgium  \and
Max-Planck-Institut f\"ur Astrophysik, Karl Schwarzschild Str. 1, 85741 Garching, Germany
}

\date{Received 20 December 2017/ Accepted 3 March 2018}

\abstract
{Most super AGB stars are expected to end their life as oxygen-neon white dwarfs rather than electron capture supernovae (ECSN). The reason is ascribed to the ability of the second dredge-up to significantly reduce the mass of the He core and of the efficient AGB winds to remove the stellar envelope before the degenerate core reaches the critical mass for the activation of electron capture reactions. }
{In this study, we  investigate the formation of ECSN through case A and case B mass transfer. In these scenarios, when Roche lobe overflow stops, the primary has become a helium star. With a small envelope left, the second dredge-up is prevented, potentially opening new paths to ECSN. }
{We compute binary models using our stellar evolution code {\sf BINSTAR}. We consider three different secondary masses of 8, 9, and 10\Msun{} and explore the parameter space, varying the companion mass, orbital period, and input physics.}
{Assuming conservative mass transfer, with our choice of secondary masses all case A systems enter contact either during the main sequence or as a consequence of reversed mass transfer when the secondary overtakes  its companion during core helium burning. Case B systems are able to produce ECSN progenitors in a relatively small range of periods ($3\la P(\mathrm{d})\le 30$) and primary masses ($10.9 \le M/M_\odot\le 11.5$). Changing the companion mass has little impact on the primary's fate as long as the mass ratio $M_1/M_2$ remains less than $1.4-1.5$, above which evolution to contact becomes unavoidable. We also find that allowing for systemic mass loss substantially increases the period interval over which ECSN can occur. This change in the binary physics does not however affect the primary mass range. We finally stress that the formation of ECSN progenitors through case A and B mass transfer is very sensitive to adopted binary and stellar physics.}
{Close binaries provide additional channels for ECSN but the parameter space is rather constrained likely making ECSN a rare event.}

\keywords{stars:binaries -- white dwarfs -- supernovae:general }

\titlerunning{}
\authorrunning{L. Siess }

\maketitle

\section{Introduction}

In the past decade, super-AGB (SAGB) stars have generated a resurgence of interest in the stellar evolution community; they represent a non-negligible fraction of stars in the Galaxy and until recently their contribution to the galactic chemical enrichment was largely ignored, mainly due to the lack of reliable yields. Their peculiar evolution and in particular their fate as electron capture supernovae (ECSN) is also a matter of active research. The final evolution of SAGB stars still represents a challenge to one-dimensional (1D) modelling with off-center neon ignition and/or silicon burning flame that propagate to the center \citep{Woosley2015}, possibly leading in some cases to the disruption of the SAGB core and the formation of iron remnants \citep{Jones2016}. Hydrodynamical simulations of ECSN \citep[for a review, see][]{Muller2016} provide successful explosions after core bounce and subsequent neutrino heating. These supernovae produce the lowest mass neutron  stars and because the explosion is fast and not very powerful ($\approx 10^{50}$~erg), the neutron star is expected to receive a small natal velocity kick. This weak impulse is invoked to explain the neutron star retention problem in globular clusters \citep{Pfahl2002} or the small eccentricity of some high-mass X-ray binaries such as X Per \citep{Pfahl2002b}. We note however that there is no clear evidence for a bimodal velocity distribution in pulsars kicks \citep{Hobbs2005} that would identify the neutron star explosion mechanism as EC or core collapse (CC) supernova.

SAGB stars are members of a specific class of stars squeezed between intermediate-mass stars that end their lives as CO white dwarfs (WD) and massive stars that form neutron stars or black holes after they explode as core collapse supernova (CCSN). The mass of SAGB stars ranges between $M_\mathrm{up}$ the minimum mass for carbon ignition and $M_\mathrm{mas}$ the minimum mass for CCSN. The value of these dividing masses critically depends on the adopted mixing prescription at the edge of the convective core and varies
between $M_\mathrm{up} \approx 6 - 8$\Msun{} and $M_\mathrm{mas} \approx 10-12$\Msun{} \citep[e.g.,][]{Siess2006}. The evolution of SAGB stars  \citep[for a recent review, see][]{Doherty2017} is characterized by the off-center ignition of carbon followed by the propagation of a deflagration front to the center and the formation of an oxygen-neon core. The subsequent evolution depends on the ability of the degenerate core to reach the critical mass of $M_\mathrm{EC} = 1.37$\Msun{} \citep{Nomoto1984} above which the density is high enough for electron capture reactions on the abundant \chem{20}Ne to take place. If this threshold is reached, the reduction of the electron number induces the collapse of the stellar core and the formation of a low-mass neutron star. On the other hand, if the SAGB star is able to get rid of its
envelope because of efficient wind mass loss for instance, the end product of the evolution is an ONe WD. The critical  mass that delineates these two fates is usually referred to as $M_\mathrm{n}$, the
minimum mass for the formation of a neutron star.

Results of full stellar evolution calculations indicate that the mass range of single stars that undergo an ECSN is very narrow, with $M_\mathrm{mas}-M_\mathrm{n} \approx 0.1-0.3$\Msun{} \citep{Doherty2015}. There are two main reasons why so few SAGB stars follow this explosive path. The first one has to do with the occurrence of the second dredge-up (2DUP). Indeed, at the end of core helium burning, the expansion of the star to red giant dimensions is accompanied with the deepening of its convective envelope. In SAGB and massive intermediate-mass stars, the surface convection zone reaches the outer He-rich layers and
reduces the He core mass below the Chandrasekhar limit \cite[see e.g. Fig 5 of][]{Siess2006}. It is worth reiterating that massive stars do not experience a 2DUP event and maintain a massive He core that can subsequently evolve through all the nuclear burning stages until the formation of an iron core. Second, the mass loss rate during the thermally pulsating SAGB phase is so strong compared to the core growth rate that the entire SAGB envelope is lost before the core mass can reach \Mec. Therefore only stars that enter the SAGB phase after the 2DUP with a core mass close to \Mec{} may eventually go SN, the large majority ending as ONe WD. The apparent failure of single stars to evolve toward ECSN should however be mitigated because a large fraction of stars are in binary systems \citep{Raghavan2010,Duchene2013} and will interact with their companion at some point during their evolution \citep{Sana2012}. Among these interacting systems, some ECSN progenitors may emerge.

In this paper, we discuss these binary channels and more specifically those resulting from the evolution through case A and case B mass transfer. The following section sets the stage and reviews the binary paths leading to ECSN (Sect.~\ref{Sect:review}). After a description of our code and physical assumptions (Sect.~\ref{Sect:code}), we present the results of our simulations, starting with a description of the evolution of representative case A and case B systems. We then explore the effect of varying the initial
period, the mass ratio\footnote{We define the mass ratio as the ratio of the actual donor's mass to that of the gainer: $ q = M_1/M_2$.} and the assumptions concerning conservative mass transfer (Sect.~\ref{Sect:results}). In Sect.~\ref{Sect:Hestar}, we analyze the mass of the He star after case B Roche lobe overflow (RLOF) and of the envelope at the pre-ECSN stage. In Sect.~\ref{Sect:discussion}, we discuss the uncertainties affecting this modeling and compare our results with the recent work of \cite{Poelarends2017} before concluding in Sect~\ref{Sect:conclusion}.

\section{The binary channels to ECSN}
\label{Sect:review}

Single SAGB stars fail to explode as ECSN because their ONe core mass cannot reach the critical value of 1.37\Msun. One way to overcome this problem is to increase the core mass by adding matter on top
of it. Accretion from a binary companion is an obvious means to do this, but depending on the composition of the accreted material, different scenarios have to be considered: either a CV-like evolution in the case of accretion of H-rich material (Sect.~\ref{sect:AIC}) or the merger of two WDs when the accreted matter is made of C and O (Sect.~\ref{sect:merger}). In Sect.~\ref{sect:Hestar} we describe the evolution of helium stars whose characteristics are similar to those of the bare core of SAGB stars except that their envelope has been removed.

\begin{figure}
\centering
\includegraphics[width=\columnwidth]{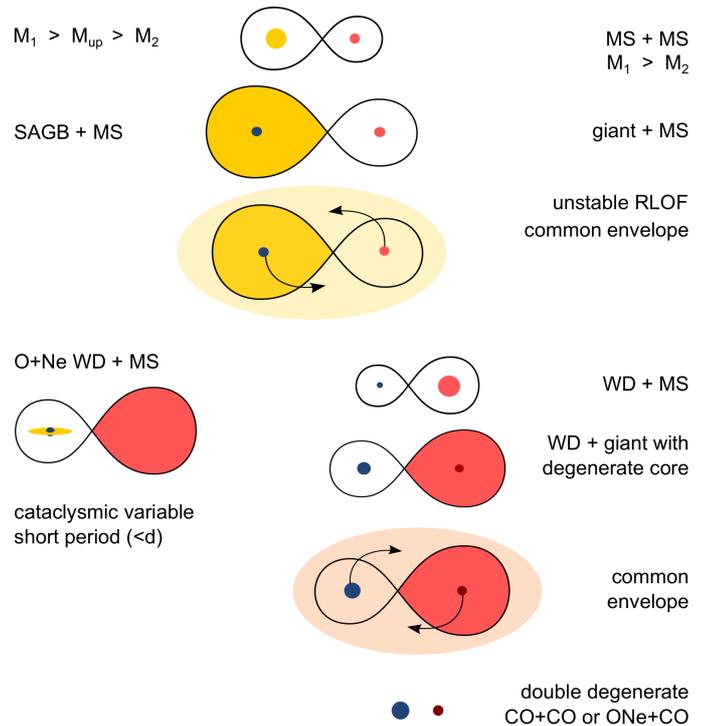}
\caption{Cartoon illustrating the formation of an EC progenitor in a single (left) and double degenerate (right) scenario.}
\label{fig1}
\end{figure}

\subsection{Accretion-induced scenario}
\label{sect:AIC}

Short period binaries consisting of an O+Ne WD accreting material from a main sequence companion represent the high-mass analogs of cataclysmic variables (CV) and the formation of these systems is similar to that of CVs. In this paradigm \citep[]{Ritter2012}, a SAGB primary fills its Roche lobe as it ascends its thermally pulsing SAGB phase (Fig.~\ref{fig1}, left). Because the donor star has an extended convective envelope, it expands upon mass loss and mass transfer becomes dynamically unstable\footnote{This requires however that at the start of Roche lobe overflow the mass ratio is large enough \cite[$q=M_1/M_2 \ga 1.2-1.5$, see][]{Webbink1988}}. In this process, the low-mass companion cannot assimilate the matter that is rapidly dumped on its surface and the binary becomes immersed in a common envelope. During the subsequent evolution ($\sim 1000$\,yr), the stars spiral towards each other as angular momentum and potential energy are transferred by friction from the orbit to the envelope \citep[for a review on common envelope evolution,   see e.g.,][]{Ivanova2013}. The outcome depends on the initial conditions (and on the assumptions concerning the efficiency of envelope ejection) and can either lead to coalescence or to the formation of a close binary system.

The existence of short period binaries hosting a ONe WD and a lower-mass main sequence companion has been attested by observations of neon novae \cite[e.g.,][]{Starrfield1986,Downen2013}. The fate of the WD is then mainly dictated by the mass accretion rate \citep{1982ApJ...253..798N,2007ApJ...663.1269N}: if the mass transfer is higher than the core growth rate then the H-rich material cannot be assimilated by the WD and expands to red-giant dimensions, potentially leading to a CE evolution. On the other hand, if it is too low, the H-shell becomes unstable and recurrent novae-like flashes develop that remove most of the accreted mass, thus preventing core growth. In the intermediate regime, steady H-burning takes place and the core mass can increase. When the WD reaches \Mec, EC reactions start on \chem{24}Mg and when the central density $\rho_c \approx 9\times 10^9$~\gcm{} they proceed on the abundant \chem{20}Ne \citep{Nomoto1984,Nomoto1987}. These reactions induce a rapid contraction of the core, which is accelerated by the strong dependence of the EC rates on density, and the rise of the temperature by the emission of $\gamma$-rays. When the central temperature $T_c \approx 2\times 10^9$~K, oxygen burning starts,  but the outcome of the evolution (collapse to a neutron star or core disruption) depends on the density $\rho_\mathrm{ig}$ at which oxygen is ignited. If the density is high enough, $\rho_\mathrm{ig}  \ga 9\times 10^9$~\gcm, EC on the NSE material left behind by the passage of the oxygen deflagration is fast enough to induce gravitational collapse before the expansion induced by thermonuclear energy release quenches them \citep{Isern1991,Canal1992,Gutierrez1996}.

As reported by various authors, many uncertainties in the adopted physics impact $\rho_\mathrm{ig}$ and thus the outcome \citep[e.g.,][]{Isern1994}. Among them is the critical treatment of mixing \citep{Gutierrez1996}. In the early stage of collapse, semi-convection develops at the center.  Using the Ledoux instead of the Schwarzschild criterion would then lead to a less efficient cooling of the central region and thus to a higher central temperature and lower density at the time of oxygen ignition. \cite{Isern1991} also showed that if the velocity of the deflagration front, which governs the nuclear energy production, exceeds some critical value (which depends on the mode of energy transport, conduction or convection), then EC reactions on the NSE material behind the oxygen deflagration front may not be fast enough to induce collapse. A recent study by \cite{Schwab2015} confirms the absence of convection at the time of activation of the EC reactions implying low ignition densities ($\rho_\mathrm{ig} \ga 8.5\times 10^9$\gcm) but the authors also conclude
that at this density, the star should still collapse to a neutron star. Using the model of \cite{Schwab2015}  as initial conditions for their 3D hydrodynamical simulations, \cite{Jones2016} find on the contrary that the core is partially disrupted. Explosive oxygen burning provokes the ejection of a fraction of the WD material and leads to the formation of a bound remnant composed of O-Ne and Fe-group elements \citep[see also][]{Isern1991}. On the other hand, if a neutron star is formed, the result of this AIC scenario may be the formation of a millisecond pulsar \citep{Tauris2013_558}.

\subsection{The merger scenario}
\label{sect:merger}

The formation of a double degenerate system is illustrated in Fig.~\ref{fig1} (right) and involves two common envelope phases.  The merger of two WD was first modeled using 1D stellar evolution codes by \cite{Nomoto1985} and \cite{Saio1985}. In this scenario, examined earlier by \cite{Iben1984} and \cite{Webbink1984}, the system comes into contact due to the loss of angular momentum by gravitational wave emission. The least massive WD, which has the larger radius, overfills its Roche lobe and is subsequently tidally disrupted. The matter is then assumed to distribute in a disk allowing the deposition of CO rich material on the more massive WD. The authors showed that if the mass accretion rate is high enough (higher than 1/5 of the Eddington limit, $\sim 4\times 10^{-6}$\myr), then carbon ignites off-center and is followed by the propagation of a burning front that converts the entire CO core in an ONe mixture,  like in SAGB stars. If the mass of
the merger exceeds \Mec, then an ECSN is a likely outcome. On the other end, if the mass accretion rate is too small, the inner shells heat faster than the surface layers and the central ignition of carbon at higher density leads to a SNIa explosion. During the merger process, the WD may be significantly spun up  but the effects of stellar rotation have been shown to be small \citep{Saio2004}.

This classical picture has recently been contested in light of hydrodynamical calculations of WD merger events. These simulations \citep[e.g.,][]{Shen2012,Raskin2012} show that once the lower-mass WD fills its Roche lobe, mass transfer becomes unstable and leads to the complete disruption of the star. At the end of this dynamical phase that lasts $\sim 10^2-10^3$s, the accretor is surrounded by a fast rotating hot envelope attached to a centrifugally supported thick disk. A stream of matter is also present in the simulations but the material remains bound to the system and eventually falls back onto the accretor. \cite{Shen2012} simulations also indicate that after the merger product has reached a quasi-hydrostatic equilibrium configuration, magneto-rotational instabilities develop in the disrupted WD material and efficiently redistribute the angular momentum. According to their simulations, within $10^4-10^8$s, the remnant evolves towards an equilibrium configuration of shear-free solid body rotation. In this process, viscous heating has substantially raised the temperature of the envelope, potentially allowing for C ignition already during the merger phase. During this ``viscous-phase'', the merger product reaches a nearly spherical geometry, with a cool CO WD at the center, surrounded by a thermally supported envelope \citep{Schwab2012}. These conditions are quite different from those used in earlier studies where the primary was accreting mass at a nearly Eddington rate from a keplerian disk. In this new configuration, there is no accretion and the evolution of the remnant is determined by the cooling and heat redistribution in the hot envelope. \cite{Yoon2007} were the first to implement the results of a 3D SPH simulation of a WD merger into a 1D stellar evolution code. However, they did not consider the effect of MHD instabilities on the redistribution of angular momentum, so their starting structure is slightly different, consisting of  a cool WD surrounded by a hot envelope gaining mass from an accretion disk. In their analysis, they derive the conditions for off-center carbon ignition and show that they depend on the initial temperature, mass accretion rate, and efficiency of angular momentum transport. Using more realistic initial conditions that take into account the viscous phase, \cite{Schwab2016} followed the evolution of a 0.9+0.6\Msun{} merger. In their model carbon ignites off-center during the dynamical phase. The propagation of a carbon burning front towards the center is similar to that occurring in SAGB stars and results in the formation of a ONe core. However, when the carbon flame reaches the center, the degeneracy has been lifted. The core then contracts, neon ignites off-center and a Ne-O deflagration propagates to the center converting the core into a hot mixture of Si-group elements. The final outcome depends on the core mass: if it is less than $\sim 1.41$\Msun, the remnant is a silicon WD, otherwise silicon burning proceeds and the core collapses to a neutron star. The authors suggest that the same conclusion could have been reached by \cite{Nomoto1985} if they had been able to run their simulations for a longer time. These results depend, however, on the efficiency of mass loss during the heat redistribution phase but the important point is that CO WD mergers may not provide a viable path toward ECSN.

For the merger of a ONe and CO WDs, \cite{Kawai1987} showed that the compressional heating induced by mass accretion at a nearly Eddington rate is not strong enough to ignite neon off-center. In this case, once a sufficient amount of carbon has been accreted, carbon burning operates above the O+Ne core and contributes to increase the WD mass. The process of carbon accretion/burning repeats until the WD mass reaches \Mec{} and EC reactions begin.

\begin{figure}
\centering
\includegraphics[width=\columnwidth]{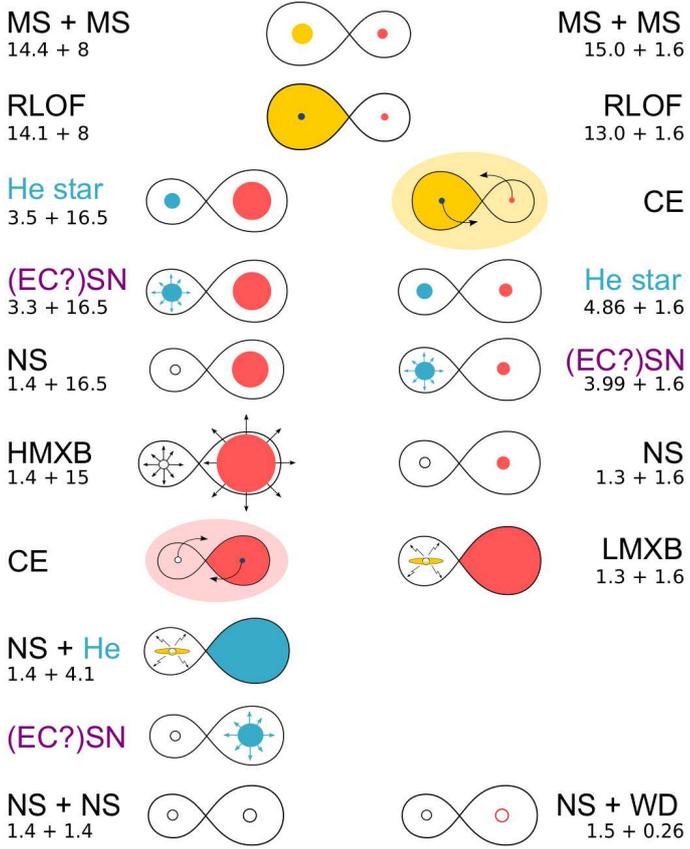}
\caption{Cartoon illustrating the possible occurrence of ECSN after the formation of a He star in the scenarios involved for the formation of high- (left) and low- (right) mass X-ray binaries. This cartoon has been adapted from \cite{Tauris2006}. The numbers indicate the masses of the stellar components.}
\label{fig2}
\end{figure}

\subsection{The helium star progenitors}
\label{sect:Hestar}

In the early 1980's, \cite{Nomoto1984,Nomoto1987} and \cite{Habets1986} studied the evolution of He stars and showed that models with initial masses ranging between 2.0\Msun{} and 2.5\Msun{} would evolve toward ECSN.  But these single stellar models do not take into account the effects of binary interactions that can deeply impact the progenitor's structure. Indeed, the final He core mass depends on the presence of a convective envelope because of the contribution coming from the ashes of H-shell burning. The models of \cite{Wellstein1999} show that a donor star that loses its envelope will develop a smaller He core compared to a single evolution.  In addition, the exchange of mass and angular momentum between the binary components can induce hydrodynamical instabilities responsible for the transport of chemicals and angular momentum. Rotational mixing can strongly affect the size of the stellar core \citep[e.g.,][]{Maeder2000} and the entire structure up to the point that, if the system is near contact, the stars may be tidally locked and follow a chemically homogeneous evolution \citep[e.g.,][]{deMink2009}.

\cite{Podsi2004} also pointed out that if the star is able to lose its H-rich envelope by the end of core He-burning, the second dredge-up can be avoided, thereby removing one of the main factors preventing the evolution toward ECSN. Using \cite{Nomoto1984} He core mass range and binary models from \cite{Wellstein2001}, \cite{Podsi2004} estimated that stars in a binary system with masses in the range $8-11$\Msun{} would likely undergo an ECSN. As we will show, consistent binary models considerably reduce this primary mass range. The evolution of naked stellar cores has been the subject of various investigations, largely related to the formation of low- and high-mass X-ray binaries as well as millisecond pulsars \cite[see e.g., the review by][]{Tauris2006}. Helium stars can form via common envelope evolution or as a result of case A and case B mass transfer and are thus a very common outcome of binary evolution.

In a recent work, \cite{Tauris2015} analyzed the evolution of close binary systems composed of a He star donor and a neutron star companion with initial periods between 0.06 and 2 days. This period range was chosen so the primary would fill its Roche lobe during its evolution. Their simulations indicate that ECSNs occur over a narrow mass range of $\approx 0.1-0.2$\Msun{} and involve He stars with masses between 2.6\Msun{} and 2.95\Msun{} (see their Fig.~18).

\section{Modeling and Methodology}
\label{Sect:code}

\subsection{Input physics and binary modeling}
\label{Sect:physics}

The calculations presented in this paper have been performed with the {\sf BINSTAR} code whose detailed description can be found in \cite{Siess2013} and \cite{Davis2013}. In brief, the code solves the structure of the two stars and the evolution of the orbital parameters (eccentricity and separation) simultaneously. When the star fills its Roche lobe, mass transfer rate is calculated according to the prescriptions of \cite{Ritter1988} and \cite{Kolb1990}, updated at each iteration during the convergence process. The Roche radius is approximated by the \cite{Eggleton1983} formula. In our simulations we consider circular orbits and neglect the stellar spins. The exchange of mass between the stellar components is accompanied by the transfer of angular momentum. If the evolution is not conservative, we assume that the material leaving the system carries away the specific angular momentum of the gainer star at its position along the orbit (so-called re-emission mode). In all our models, we use the \cite{Asplund2005} solar composition, neglect stellar winds, and use a mixing length parameter $\alpha = 1.75$.  We apply a moderate core overshooting modeled with the use of an exponentially decaying diffusion coefficient outside the Schwarzschild boundary with a parameter $f_\mathrm{over}=0.01$ \cite[for details of this implementation in {\sf STAREVOL}, see][]{Siess2007}. Our nuclear network includes all the necessary reactions to accurately follow the energetics up to neon ignition and our simulations are stopped once the maximum temperature reaches $\sim 1.5\times 10^9$K. With our assumptions, we find that single stars in the mass range  $9.7M_\odot \la M_\mathrm{zams} \la 9.9M_\odot$ go ECSN.

\begin{figure}
\centering
\includegraphics[width=\columnwidth]{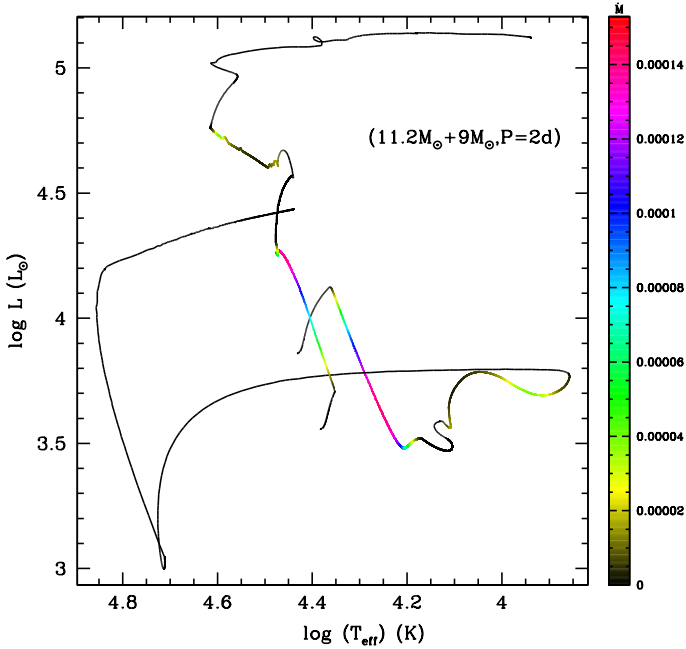}
\caption{HR diagram of a representative 11.2+9\Msun, two-day case A evolution. The color bar on
  the side indicates the magnitude of the mass transfer rate.}
\label{fig:HR_caseA}
\end{figure}

\subsection{Determining the fate of the system}
\label{Sect:criteria}

With our limited nuclear network, we are not able to follow the evolution up to the final stage. Therefore, we had to devise criteria to determine the most likely fate of our stellar models. Our method relies on several indicators. The first systematic one is borrowed from \cite{Tauris2015} and declares a model as ECSN progenitor if, during the post-carbon burning phase, the central temperature $T_c$ does not exceed the maximum value reached during the core carbon-burning phase. According to these authors this corresponds to stars with an ONe core mass in the range 1.37\Msun{} to 1.43\Msun. Therefore any star that ends up with a CO core mass in that range and fulfills the temperature condition is declared a ECSN progenitor. The second indicator is mostly a check based on a visual inspection of the evolution of the star in the central density  versus central temperature ($\rho_c-T_c$) diagram. We simply make sure that for the selected ECSN progenitor, the increase in the central temperature  after core carbon burning is modest. Stars that evolve towards an ONe WD show a pronounced decrease in their central temperature while in massive stars the temperature rises steadily with increasing density.

\section{Results}
\label{Sect:results}

Before exploring the parameter space leading to ECSN, we start our study with the analysis of two representative systems illustrating a conservative evolution during case A and case B evolution.

\begin{figure}
\centering
\includegraphics[width=\columnwidth]{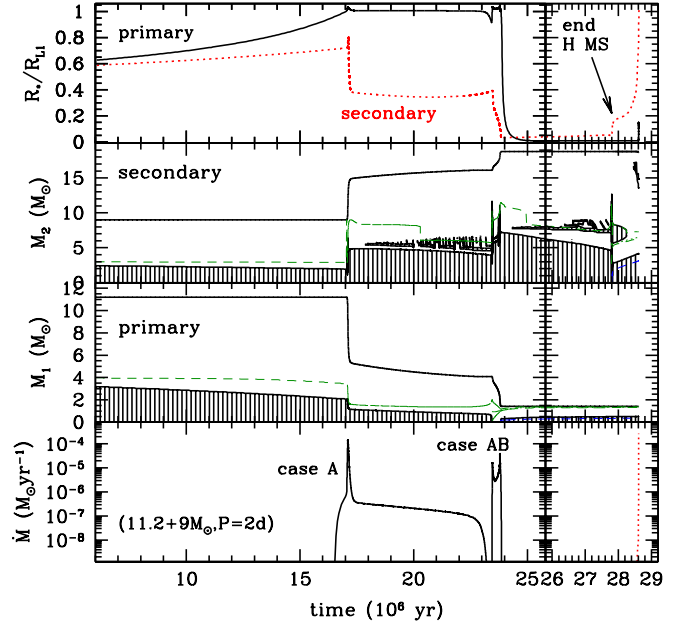}
\caption{Evolutionary properties of a representative  11.2+9\Msun, two-day case A system. From top to bottom, the panels show the overfilling factor (ratio of the stellar to Roche radius), the Kippenhahn diagrams of the secondary and of the primary, and the Roche lobe overflow mass transfer rate. The red dotted lines in the top and bottom panels refer to the secondary.}
\label{fig:caseA}
\end{figure}

\subsection{Case A systems}
\label{sect:caseA}

\begin{figure}
\centering
\includegraphics[width=\columnwidth]{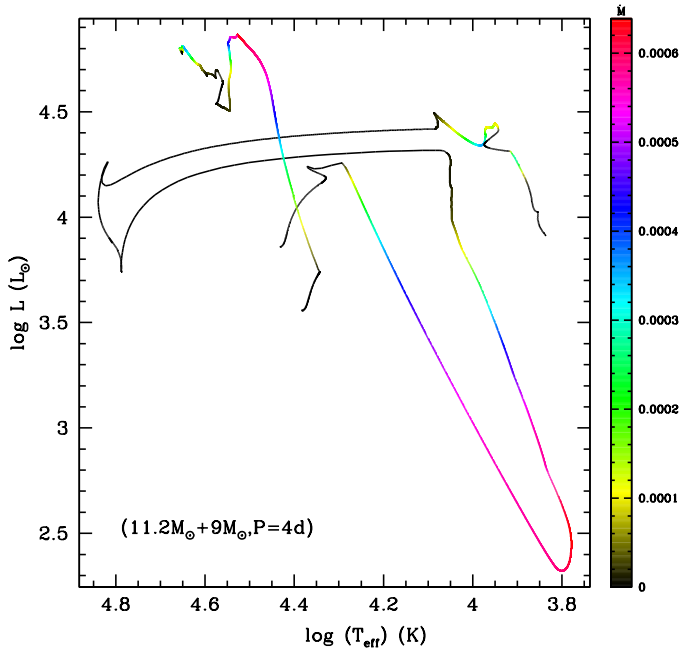}
\caption{HR diagram of a representative 11.2+9\Msun, four-day case B evolution. The color bar on the side indicates the magnitude of the mass transfer rate.}
\label{fig:HR_caseB}
\end{figure}

Our prototype is a 11.2\Msun{} primary with a 9\Msun{} companion in an initial period of 2 days.  The evolution of the system in the HR diagram is presented in Fig.~\ref{fig:HR_caseA}. The primary starts filling its Roche lobe while on the main sequence. The mass transfer (Fig.~\ref{fig:caseA}) shows the typical pattern with a rapid, thermally unstable phase during which most of the mass is transferred followed by a slower phase where the mass transfer is driven by the nuclear expansion of the star on the main sequence. The system temporarily detaches with the exhaustion of nuclear fuel in the core but mass transfer resumes soon after the establishment of H-shell burning, an episode referred to as case AB. The system detaches when the primary contracts as a result of central helium ignition. At that stage, the secondary has gained $\approx 10$\Msun{} and the binary is now composed of a 1.4\Msun{} He star primary and a 18.8\Msun{} main sequence companion orbiting each other with a period of $\approx 110$~d. We also note in the Kippenhahn diagram of Fig.~\ref{fig:caseA} the rejuvenation \citep[e.g.,][]{DeGreve1990} of the secondary characterized by the increase of its convective core upon accretion. Because of its high mass, the secondary now evolves much faster than the primary and is able to overtake its companion during core helium burning. The secondary leaves the He main sequence first and the subsequent expansion of its radius leads to a new episode of mass transfer. However given the extreme mass ratio ($M_2/M_1 > 13$) of this system, a common envelope (CE) evolution is unavoidable. This {\em reverse case C } evolution is found in all our conservative case A evolution (see \ref{sect:EC_channel}) and corresponds to the {\em late overtaking} type described in \cite{Nelson2001}.
To guess the outcome of this system, we followed \cite{Dewi2000} and computed the expected final separation assuming that the Roche overfilling secondary would lose all its H-rich envelope, technically defined as the mass coordinate where the H mass fraction drops below 0.1. Using a CE efficiency parameter $\eta_\mathrm{CE} = 1$ and the binding energy computed with our stellar models, we find that at the end of the CE evolution the secondary still fills its Roche lobe ($R_{L1}/R_2 \approx 0.25$) so a merger is a likely outcome. However if we approximate the binding energy using Eq.~4 of \cite{Dewi2000} with $\eta_\mathrm{CE}\lambda = 1$, we find that $R_{L1}/R_2 \approx 1.8$, so in this case a merger is avoided. It is therefore difficult to conclude on the fate of the binary. We extended our analysis to other reverse case C systems and reached the same conclusion because all these binaries share the same properties (periods between 80 and 120~d, $M_1$ around 18\Msun, $M_2$ of
  the order of $\sim 1.5-1.8$\Msun{} and radius $R_2 \approx 0.4-0.7$\Rsun).

We conclude that with our choice of secondary masses, case A systems fail to produce ECSN progenitors.
A successful evolutionary path requires a more massive primary that can develop a bigger He core and evolve rapidly enough to avoid being overtaken by its companion  \citep[for details see][]{Poelarends2017}.

\subsection{Case B systems}
\label{sect:caseB}

\begin{figure}
\centering
\includegraphics[width=\columnwidth]{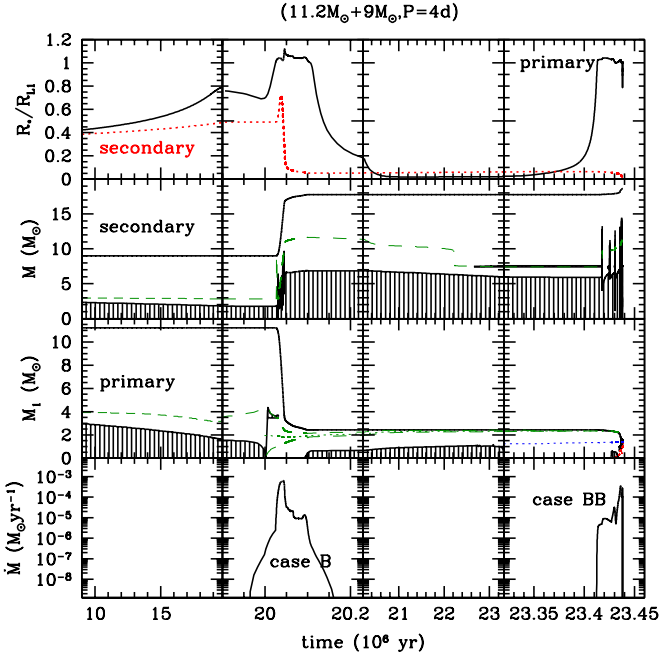}
\caption{Representative evolution of a 11.2+9\Msun, four-day case B system. From top to bottom: overfilling factor (ratio of the stellar to Roche radius), Kippenhahn diagrams of the secondary and of the primary and Roche lobe overflow mass transfer rate. The red dotted lines in the top panel refer to the secondary.}
\label{fig:caseB}
\end{figure}

\begin{figure}
\centering
\includegraphics[width=\columnwidth]{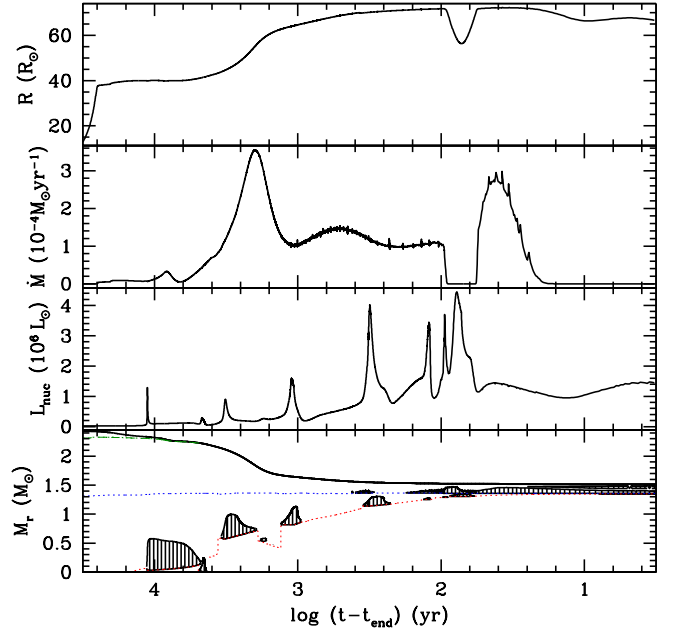}
\caption{As in Fig.~\ref{fig:caseB} but zooming into case BB. Time is counted backward from the last computed model. Mass transfer stops around log$(\mathrm{t-t_{end}}) \approx 1.2$.}
\label{fig:caseBB}
\end{figure}

To illustrate case B evolution, we consider the same system but with a longer initial period of 4 days. The evolution in the HR diagram is presented in Fig.~\ref{fig:HR_caseB}. With larger initial separation, mass transfer begins after the primary has left the main sequence and proceeds on its thermal timescale. The donor star being more evolved than in case A, its Kelvin Helmholtz timescale is shorter leading to a higher mass transfer rate with a maximum value of $\sim 7\times 10^{-4}$\myr{} compared to $\sim 10^{-4}$\myr{} in case A (Fig.~\ref{fig:caseB}). In this scenario, the slow phase is absent and when the system detaches (after $\sim 6\times 10^4$yr), the primary, now a He star, is substantially more massive than in case A with a mass of 2.5\Msun{} and the period is shorter with $P\approx50$~d. The He star has a small H-envelope of $< 0.1$\Msun{} and terminates central He-burning while its companion is still on the main sequence. The exhaustion of fuel causes the expansion of the primary and triggers a new episode of mass transfer, referred to as case BB. The mass transfer rate is slightly less intense than before ($\dot{M} < 4\times 10^{-4}$\myr) and stops after $\sim$ 20500~yr when neon ignites off-center. As illustrated in Fig.~\ref{fig:caseBB}, the mass transfer rate during case BB presents strong variations associated to the appearance of carbon burning shells, a feature also reported by \cite{Dewi2002}. Since the H-rich envelope has been almost entirely removed, mass transfer starts to erode the He core acting like the second dredge-up in a single SAGB star. However, the system detaches before the core mass drops below the Chandrasekhar limit, leaving a good candidate for an ECSN with a CO core mass of 1.38\Msun.  At the end of the simulation, the period is $\sim 180$~yr and the companion is now a main sequence O star of 18.6\Msun. After the SN explosion of the He star, the system will likely appear as a high mass X-ray binary where the neutron star accretes matter from the companion's wind.

\subsection{Dependence of the primary's evolution on the initial period}
\label{sect:EC_channel}

To analyze how the fate of the primary depends on the initial period, we consider a system with a 9\Msun{} companion. In the following section, we investigate the effect of changing the mass of the secondary.

\begin{figure}
\includegraphics[width=\linewidth]{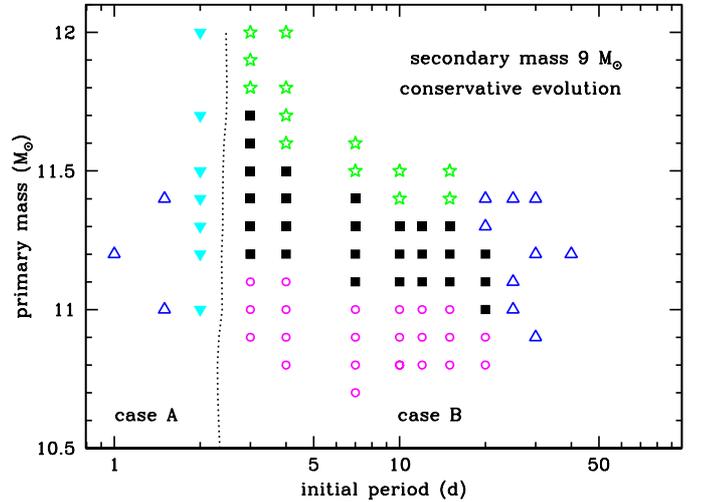}
\caption{Fate of the primary star as a function of its mass and initial period for a system with a 9\Msun{} secondary companion. The triangles indicate an evolution to contact with the cyan triangles corresponding to systems that undergo a reverse case C mass transfer. The open magenta circles and black filled squares label binaries in which the primary end its life as an ONe WD and ECSN, respectively. The green stars identify systems with a massive He core that evolves as CCSN. The transition between case A and case B mass transfer is delineated by the dotted line and occurs for periods between 2.3 and 2.5 days.}
\label{fig:period9}
\end{figure}

The results of our binary simulations are presented in Fig.~\ref{fig:period9}. Systems with periods shorter than $\approx 2.4$~d  undergo case A mass transfer and evolve into contact either while the stars are on the main sequence (MS) or as explained in $\S$~\ref{sect:caseA} when the secondary overtakes the evolution of the primary and initiates a reversed mass transfer when He-shell burning begins. Contacts on the MS come from different origins. For very short periods $P\la 1$~day (e.g., a 11+9\Msun, 1 d system), the secondary is initially close to filling its Roche lobe and contact occurs during the slow nuclear timescale mass transfer (case AS in \cite{Nelson2001} terminology). With increasing separation (e.g., a 11+9\Msun, 1.5~d system) the companion is deeper inside its potential well and accepts the accreted matter without overfilling its Roche lobe. However, its evolution is accelerated to the extent that it leaves the MS before the donor star. This situation corresponds to the early overtaking or {\em premature contact} described by \cite{Wellstein2001}. These authors also emphasized that the occurrence of contacts depends on the criterion used to define convection. In particular, using the Ledoux criterion tends to suppress rejuvenation \citep{Braun1995}, resulting in secondaries with smaller cores that have a shorter main sequence lifetimes and are consequently more prone to premature contact. \cite{Pols1994} analyzed the evolution of close binaries with donor stars in the mass range $8-16$\Msun{} and found that case A systems with a mass ratio $q \ga 1.4-1.5$ evolve into contact during the fast mass transfer phase. This result was corroborated by \cite{Wellstein2001} and is confirmed by our models.

For our considered range of stellar masses, the transition between case A and case B occurs between 2.3 and 2.5~d. In the period range $3\la P (\mathrm{d}) \la 20$, we see in Fig.~\ref{fig:period9} that a variety of outcomes become possible. Let us analyze for instance systems with an initial period $P=4$~d. For a donor star with $M_1 < 11.1$\Msun, the He-star left at the end of RLOF has a mass less than $\approx 2.5$\Msun{} (see Sect.~\ref{Sect:Hestar}). In these models, carbon ignites off-center and leads to the formation of an ONe core less massive than \Mec. The primary thus ends up as a degenerate ONe WD. Between $11.2 \la M_1/M_\odot \le 11.5$, the ONe core fulfills the criteria defined in Sect.~\ref{Sect:criteria} and the star evolves towards ECSN. Above $M_1 > 11.5$\Msun, at the end of core carbon burning, the ONe core exceed 1.43\Msun, Ne ignites and the evolution proceeds to CCSN. It is also expected that if the mass ratio becomes even larger ($q > 1.5$), the system
  goes into contact during the fist rapid mass transfer phase \citep{Pols1994}.

With increasing period, the primary fills its Roche lobe at a more advanced stage of its evolution. It has a shorter Kelvin Helmholtz timescale ($\tau_\mathrm{KH}$) and may have developed a deep convective envelope. Since the maximum mass transfer rate scales as $\dot{M} \approx M_1/\tau_\mathrm{KH}$, these late case B systems will come into contact and enter a CE evolution. This transition to contact occurs with initial periods around $P \approx 20-25$~d. We reiterated the same procedure described at the end of Sect.~\ref{sect:caseA} to estimate the final separation after the CE evolution. At the time of contact, our late case B systems typically consist of a H-shell burning donor of 5-6\Msun{} with a He core of $1.6-1.7$\Msun{} and a main sequence companion of $14-15$\Msun. With periods in the range $P\approx 40-70$~d, assuming that the gainer is unaffected in the process, we find that after the removal of its envelope, the core of the primary still largely overfills its Roche lobe. These systems are thus likely going to merge.

Finally, for very long initial periods, the system remains detached and one recovers the single star channel for ECSN, i.e., primaries with masses in the range $9.7\la M_1/M_\odot \la 9.9$ (Sect.~\ref{Sect:physics}). Given that the maximum radius reached by a 9.8\Msun{} star is $\approx 490$\Rsun, binary systems with companion masses of 9\Msun{} will remain detached if the initial period $P \ga 1260$~d.

Figure~\ref{fig:period9} also shows that with increasing initial period the primary mass range for ECSN  decreases slightly. This results from the fact that in wider systems, the helium core can grow bigger before RLOF starts, and therefore the primary does not need to be initially as massive to go SN. In other words, the shorter the initial period, the smaller the mass of the He star after mass transfer, so an initially more massive primary is needed to go supernova. \cite{Tauris2013} reach the same conclusion in their simulations of He star and neutron star systems.

\subsection{Dependence on the secondary mass}

\begin{figure}
\centering
\includegraphics[width=\columnwidth]{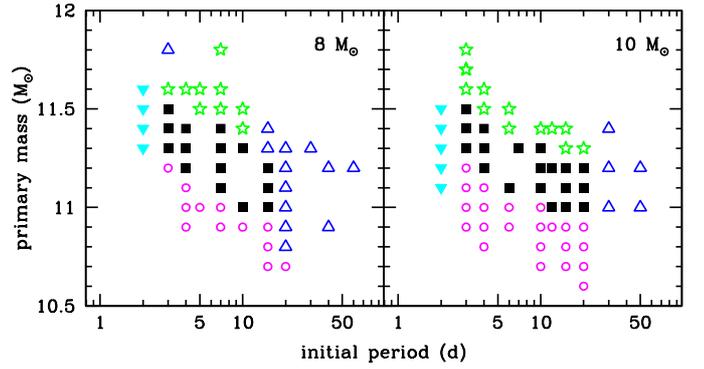}
\caption{As in Fig.~\ref{fig:period9} but for a 8\Msun{} (right panel) and 10\Msun{} (left panel) secondary.}
\label{fig:periods}
\end{figure}

The effect of varying the companion's mass is illustrated in Fig.~\ref{fig:periods}. Overall, the dependence on the secondary mass is weak because once RLOF starts, nothing can really prevent the primary's envelope from being lost. For a given initial separation, increasing the
companion mass will trigger mass transfer earlier because the primary's Roche radius, which is an increasing function of the mass ratio, is smaller. As a consequence the primary mass range for ECSN is slightly shifted downward when $M_2$ is larger.

For companion masses below $M_2\la 7$\Msun, the ECSN channel is closed because the minimum primary mass  required for an ECSN event ($M_1\ga 11M_\odot$) yields a mass ratio $q > 1.5$, which is high enough to bring the system into contact either during case A or case B evolution. With our assumptions, we did not find any ECSN progenitors with a 7\Msun{} secondary and all the systems with $q<1.5$ that avoid contact, lead to the formation of an ONe WD plus a O star companion.

\subsection{Nonconservative evolution}

In the above simulations, we assumed that mass transfer was conservative but a fraction of the mass lost by the donor may escape from the system. This nonconservatism is often advocated in case A and B mass transfer to explain for example the mass ratio distribution of Algols \citep{Deschamps2013}. To assess the impact of systemic mass loss, we use a very simple model where we assume that a constant fraction $\beta_\mathrm{RLOF} = \sfrac{1}{2}$ of the transferred mass leaves the system, carrying away the specific orbital angular momentum of the gainer star. In this so-called re-emission mode, the torque applied on the orbit writes
\begin{equation}
\dot{J}_\Sigma = \beta_\mathrm{RLOF} \dot{M}_1 a_2^2\,\Omega
\label{eq:Jdot}
,\end{equation}
where $\dot{M}_1<0$ is the mass loss rate of the donor, $a_2$ the distance between the gainer and the center of mass, and $\Omega$ the orbital angular velocity. Interestingly, with this systemic angular momentum loss prescription, the separation starts to increase as soon as the mass ratio $q\la 1.133$\footnote{It is easy to show that under the above assumptions, for $\beta_\mathrm{RLOF} = \sfrac{1}{2}$ the rate of change of the separation writes
\begin{equation}
\frac{\dot{a}}{a} = \frac{|\dot{M}_1|}{q (M_1+M_2)}\left(2+\frac{q}{2}-2q^2\right)
\label{eq:adot}
,\end{equation}
and is positive for $q <(1+\sqrt{65})/8 \approx 1.133.$}, which is earlier than in the standard case where this occurs when $q<1$.

\begin{figure}
\includegraphics[width=\columnwidth]{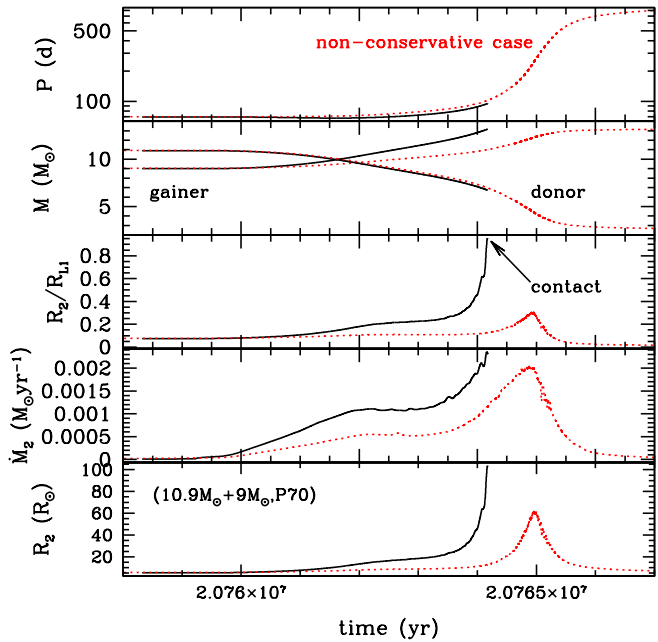}
\caption{Variation with time of selected quantities during the conservative (black, solid line) and nonconservative (red, dotted line) evolution of a 10.9+9\Msun, 70~d systems. From top to bottom are shown, the period in days, the stellar masses and the filling factor, mass accretion rate, and radius of the gainer.}
\label{fig:cons}
\end{figure}

\begin{figure}
\includegraphics[width=\columnwidth]{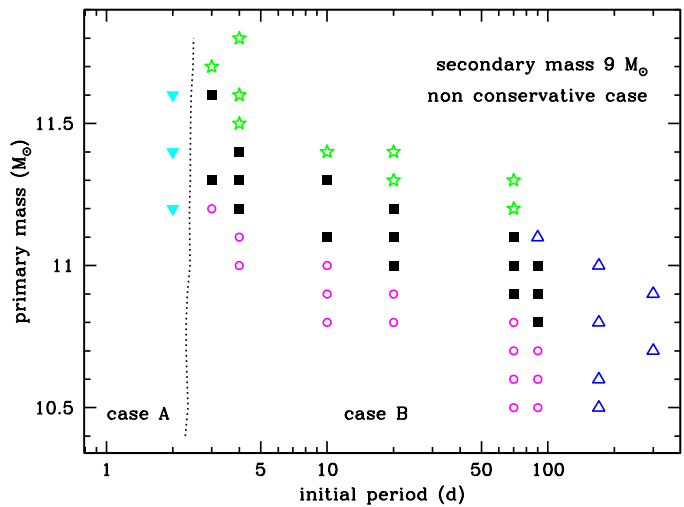}
\caption{As in Fig.~\ref{fig:period9} but for a nonconservative evolution where we assume that a constant fraction $\beta_\mathrm{RLOF} = \sfrac{1}{2}$ of the mass lost by the primary escapes from the system.}
\label{fig:beta}
\end{figure}

In a nonconservative case, the accretion timescale is a factor $\beta_\mathrm{RLOF}^{-1}$ longer than in a standard evolution. As illustrated in Fig.~\ref{fig:cons}, the expansion of the gainer is noticeably slower and much smaller when mass is allowed to escape from the system. So when the secondary reaches its maximum radius, the period has substantially increased and the star is unable to fill its Roche lobe. In our example (a 10.9+9\Msun, 70~d binary), the Roche filling factor $R_2/R_{L1} < 0.3$ and the system remains detached. With longer initial periods, the mass transfer rate is higher, the rate of expansion of the gainer becomes faster than the rate of increase of the separation and contacts become unavoidable. As shown in Fig.~\ref{fig:beta}, all late case B systems with  $P \ga 100$~d go into contact. For comparison, this threshold period is of the order of 20~d in the conservative case. The decrease in the ECSN progenitor mass with increasing period is also present in these liberal models and since the channel is now open to longer-period systems, less massive primaries can go SN but the width in progenitor's masses still remains quite narrow.

\section{He star masses}
\label{Sect:Hestar}

\begin{figure}
\includegraphics[width=\columnwidth]{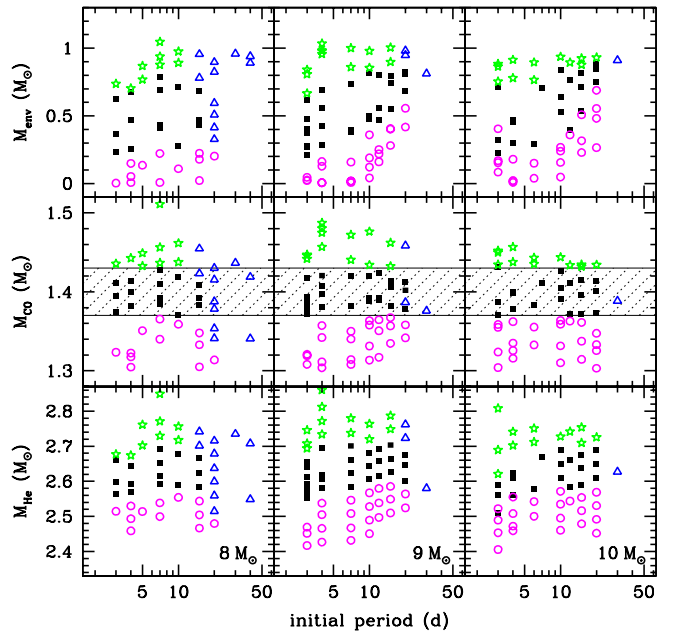}
\caption{Mass of the He star primary ($M_\mathrm{He}$) at the end of RLOF (bottom), \LS{and masses of the CO core (middle) and of the remaining envelope (top) at the end of the simulation in case B mass transfer. The final stellar mass is the sum of $M_\mathrm{CO}$ and $M_\mathrm{env}$. The hatched strip in the mid panel indicate the core mass range that we considered for the formation of ECSN progenitors.}{}  The color coding is the same as Fig.~\ref{fig:period9}.}
\label{fig:mHe}
\end{figure}

The bottom panel of Fig.~\ref{fig:mHe} shows the mass of the primary at the end of case B RLOF. All He star primaries are initially surrounded by a H envelope of a few 0.1\Msun{} but most of this H layer will be converted into He or lost during the subsequent case BB mass transfer episode. The He star mass range of ECSN progenitors is confined to a narrow region, between 2.55\Msun{} and 2.7\Msun. This result is very similar to that of \cite{Tauris2015} but different from the early models of \cite{Nomoto1984} where it was considered that He stars in the mass range $2.0-2.5$\Msun{} would undergo an ECSN.  We also see a trend of increasing He star mass with increasing orbital period. This effect is due to the occurrence of case BB mass transfer. With longer initial periods, the He star fills its Roche lobe later which implies a higher mass transfer rate and a stronger reduction of its mass before the explosion. Our simulations indicate that He stars less massive than $M_\mathrm{He} \la 2.55$\Msun{} end up as ONe white dwarfs and that above $M_\mathrm{He} \ga 2.7$\Msun{} they evolve into a CCSN. Figure \ref{fig:mHe} also shows the CO core and envelope masses at the end of our simulations. We see that our potential ECSN candidates are surrounded by a He layer of 0.1 to 0.9\Msun. With  increasing period, the  He star progenitor is more massive and the remaining envelope mass is also larger.
In this case B scenario, most of our progenitors are expected to end as SNIb since SNIc are generally attributed to progenitors with thin He envelopes \citep[$M_\mathrm{env} \la 0.06\msun$,][]{Hachinger2012} although the classification also depends on the amount of mixed \chem{56}Ni in the surface layers. We should also stress that the exact envelope mass at the time of explosion is uncertain because  in some cases mass transfer is still active and wind mass loss, which has not been considered in these simulations, may be strong in these luminous He stars.
{In terms of CO core masses, the mass interval is slightly smaller and better defined with ECSN progenitors having $2.42\la M_\mathrm{CO}/M_\odot\la 2.54$.}

\section{Discussion}
\label{Sect:discussion}

The determination of the primary mass range of ECSN progenitors is unfortunately affected by many uncertainties. The first one is associated with the treatment of core overshooting and is independent of the evolutionary scenarios. As shown in various studies \cite[e.g.,][]{Siess2007,GilPons2007,Farmer2015}, the implementation of extra-mixing at the edge of the convective core has a dramatic impact on the values of $M_\mathrm{up}$ and $M_\mathrm{mas}$ with variations up to $2-2.5$\Msun. Calculations using different levels of overshooting or different algorithms to define the convective boundaries will therefore introduce some inevitable scatter in the derived progenitors mass range.

Another issue is related to the difficulty of assessing the final evolution of some models. In some simulations, we find that at the end of carbon burning, neon ignites off-center. Similarly to off-center carbon ignition, a few neon-oxygen (NeO) shell flashes may develop before a NeO flame forms and propagates to the center. The fate of the star then depends on whether or not this flame is able to reach the center before the density reaches the critical threshold for electron capture reactions. \cite{Jones2013,Jones2014} showed that the results depend on the (unknown) mixing processes operating at the base of the convective flame. If mixing is present, the NeO flame quenches, allowing the core to contract more efficiently and reach higher densities. If instead the strict Schwarzschild criterion is used, provided neon is not ignited too far off-center, the flame can reach the center before the URCA reactions start and the star ends its life as a CCSN. Therefore the fate of these models is dictated by the mixing across the burning front which, in the absence of dedicated hydrodynamical simulations will remain a serious limitation. \cite{Jones2014} claim however that even a very small amount of mixing (of the order of $10^{-6}-10^{-5}$ times the pressure scale height at the convective front) could disrupt the flame and accelerate the contraction, in which case stars that ignite Ne off-center would never undergo an ECSN. These theoretical uncertainties have consequences on the criteria used to determine the fate of the star (Sect.~\ref{Sect:criteria}). We did a test reducing the maximum  ONe core mass of the ECSN progenitor from 1.43\Msun{} down to 1.41\Msun{} and found very small differences in the derived fate. The only  affected models are those located at the boundary between ECSN and CCSN which introduces a typical uncertainty of $\la 0.1$\Msun{} in the estimated mass range.
As outlined in the previous section, the assumptions concerning mass and angular momentum loss from the system have a deep impact on the period range over which ECSNs occur.
Various studies have shown that mass transfer in binaries is not necessarily conservative and there is no reason to assume that our ECSN binary progenitors should evolve as such. This is attested in Algols \citep[e.g.,][]{Deschamps2013,Mennekens2017} as well as in some massive binaries \citep[e.g.,][]{deMink2007,Mahy2011}. Several scenarios have been devised to account for liberality, including mass-loss enhancement due to the spin-up of the accretor \citep[e.g.,][]{Petrovic2005,Yoon2010} and the release of accretion energy in the hot spot region that forms where the accretion stream from the primary impacts the surface of the gainer \citep{vanRensbergen2008}. In our systems with periods $P\le 20$~d, mass transfer occurs via direct impact and within this latter scenario the  accretion efficiency could be significantly reduced.
Unfortunately this process is badly understood and the period range of ECSN will be subject to a large degree of uncertainty depending on the prescription used to remove mass and angular momentum from the system.

We should also bear in mind that among the systems that go into contact, in particular during case A mass transfer (the reverse case C systems indicated by the cyan triangles in Fig.~\ref{fig:period9}, \ref{fig:periods}, \ref{fig:beta}), the subsequent common envelope evolution may not necessarily lead to a merger but may produce a He star that eventually goes ECSN. Such scenarios are advocated for the formation of X-ray binaries (Fig.~\ref{fig2}) and these systems should be accounted for when estimating the total number of ECSN binary progenitors.

Recently, \cite{Poelarends2017} undertook a very similar study and a comparison of their work is instructive.
Since all our case A systems come into contact (Sect.~\ref{sect:caseA}), we focus our comparison on case B systems and on the ECSN progenitors.
A relevant difference between the two works in terms of input physics is their use of the Ledoux criterion to delineate the convective boundary and of semi-convective mixing. These assumptions will produce significantly smaller core compared to our models that consider overshooting beyond the Schwarszchild boundary. Comparing their Fig.~12 with our Figs.~\ref{fig:period9}, \ref{fig:periods} and \ref{fig:beta} indicates a systematic shift in the primary mass of $\approx 2.5$\Msun, theirs being more massive. Such differences are expected \citep[e.g.,][]{Siess2007} and do not affect the conclusion shared by the two studies that the formation of ECSN progenitors requires a minimum mass ratio of $M_2/M_1 > 0.7-0.75$. We note that taking into account wind mass loss as was done in \cite{Poelarends2017} will also contribute to favor a higher primary mass. Their period range for the occurrence of ECSN is slightly reduced compared to ours. For the conservative models, we find progenitors in the
  period interval [3,20] days while in their simulations it is restricted to [3,9] days.  The same effect is also seen in the $\beta = 0.5$ models with even bigger variations in the period range ([3,35] {\em vs.} [3,100]). The differences is likely due to the higher mass of their systems which, according to Eq.~\ref{eq:adot} induces a smaller rate of expansion of the orbit for a given $q$ and $\dot{M_1}$.
The presence of stellar winds can also contribute to reduce the orbital angular momentum. The other main difference concerns the primary mass range for ECSN progenitors which is $\approx 2$\Msun{} wide for their case B systems compared to  $\approx 0.6$\Msun{} in our study. The reason is attributed to a very efficient cooling of their structures allowing CO cores as massive as 1.52\Msun{} to go ECSN. This is substantially larger than the value we use (1.43\Msun) and that was found by \cite{Tauris2015}. It should be noted that using a pre-ECSN CO core mass of 1.43\Msun{} in their Fig.~12 considerably reduces the progenitor mass range and reconciles the two studies. In their analysis, they ascribed the efficient cooling of their models to the intense mass loss that affects the ECSN candidate in the last stage of evolution. Our models also show these high mass transfer episodes during C-shell burning (Fig.~\ref{fig:caseBB}) but they do not produce the amount of cooling the {\sf  MESA} models experience. \cite{Tauris2015} did not report this feature either. At this stage it is difficult to understand this different behavior, which may be related to a much higher mass loss rate in their simulations (they report values as high as $\dot{M} > 10^{-3}$\myr), to the equation of state, to our limited network, or to numerics (we discard the neutrino loss rate as the same prescriptions are used in both codes). \\
To conclude, the two studies are in good qualitative agreement but differences in the cooling efficiencies of the ONe core in the last stage of the evolution has a significant effect on the final mass range of ECSN progenitors.

\section{Conclusion}
\label{Sect:conclusion}

As stated above, the period and mass intervals for ECSN progenitors strongly depend on the treatment of binary interactions and stellar physics. Assuming a conservative evolution, we find that for our choice of secondary masses all case A systems enter a contact phase either when the primary is on the main sequence or when the gainer overtakes the evolution of the donor when it leaves the core helium burning phase. On the other hand, we find that case B systems with periods $3 \la P(\mathrm{d}) \la 20$ and primary masses between $10.9 \le M_1/M_\odot\le 11.5$ provide a valuable path to ECSN. Our simulations indicate that the primary mass range does not depend strongly on the companion mass provided the mass ratio does not exceed $q \la 1.5$ so contacts can be avoided. Nonconservative evolution allows longer period systems (up to $\sim 100$~d with our assumptions) and less massive primaries to go ECSN. At the end of RLOF, the mass of the He star progenitors sits in the range $2.55 \la M_\mathrm{He}/M_\odot \la 2.7$. In an early study, \cite{Podsi2004} claimed that stars in a binary system with masses between 8\Msun{} and 11\Msun{} would likely undergo an ECSN. Our consistent binary evolution calculations lead to a significant downward revision of this mass range and this conclusion is also shared by \cite{Poelarends2017}. Given the strong constraints on the parameters for stars to go ECSN, we are tempted to conclude that these explosions are rare, even including binaries. However, population studies should be performed to quantify the likelihood of these channels and investigate how the probabilities depend on the various uncertainties and scenarios.

\section*{Acknowledgments}
L.S. thanks the Max-Planck Institute for Astrophysics in Garching and in particular Achim Weiss and Ewald M\"uller for their hospitality during the final elaboration of this work. LS is senior FRS-F.N.R.S. research associate.

\bibliographystyle{aa}
\bibliography{32502}{}

\end{document}